\documentclass[twoside]{ilcws07}
\usepackage[latin1]{inputenc}
\usepackage[dvips]{graphicx,epsfig,color}
\usepackage{wrapfig,rotating}
\usepackage{amssymb,amsmath,array}
\usepackage{cite}
\usepackage{colordvi} %TR
\newcommand{\Li}[2]{{\mathrm{Li}}_{#1}(#2)}

\newcommand{\RE}{\Red}

\newcommand{\bea}{\begin{eqnarray}}%\nonumber}
\newcommand{\eea}{\end{eqnarray}}
\newcommand{\ba}{\begin{array}}%\nonumber}
\newcommand{\ea}{\end{array}}

\newcommand{\bqa}{\begin{eqnarray}}
\newcommand{\eqa}{\end{eqnarray}}
\newcommand{\nl}{\nonumber\\}

\newcommand{\eps}{{\epsilon}}

\begin{document}
\title{   %{\tiny \timestamp} ~~
%%%%   Paper title goes here  %%%%%%%%%%%%%%
New results for 5-point functions} 
%% 
%***********************************************************************
% AUTHORS INFORMATION AREA
%***********************************************************************
\author{J. Gluza$^1$ and T. Riemann$^2$%
% Optional short acknowledgment: remove next line if non-needed
\thanks{Presented by T.R.}
% DO NOT MODIFY THE FOLLOWING '\vspace' ARGUMENT
\vspace{.3cm}\\
% Addresses and institutions (remove "1- " in case of a single institution)
1 -- Institute of Physics, Univ. of
    Silesia, Universytecka 4, 40007 Katowice, Poland
%% Remove the next three lines in case of a single institution
\vspace{.1cm}\\
2 -- Deutsches Elektronen-Synchrotron DESY\\
Platanenallee 6, D--15738 Zeuthen, Germany
\\
}
%%***********************************************************************
% END OF AUTHORS INFORMATION AREA
%***********************************************************************

\maketitle

\begin{abstract}
Bhabha scattering is one of the processes at the ILC where high precision data will be expected.
The complete NNLO corrections include radiative loop corrections, with contributions from Feynman diagrams with five external legs.
We take these diagrams as an example and discuss several features of the evaluation of pentagon diagrams.
The tensor functions are usually reduced to simpler scalar functions. 
Here we study, as an alternative, the application of Mellin-Barnes representations to 5-point functions.
There is no evidence for an improved numerical evaluation of their finite, physical parts.
However, the approach gives interesting insights into the treatment of the IR-singularities.
\end{abstract}
%==========================================================================================
%==========================================================================================
%==========================================================================================
\section{Introduction}
Bhabha scattering,
\bea\label{eq1}
e^+ + e^- \to e^+ + e^-,
\eea
is one of the most important reactions at $e^+e^-$ colliders.%
\footnote{A link to the slides of this contribution is \cite{url-LCWS-loops}.
See also \cite{Riemann:2007LCWS-weak}.}
At ILC energies, small angle Bhabha scattering is dominated by pure photonic contributions and is foreseen as a luminosity monitor, and large angle Bhabha scattering is also one of the reactions with an expected jigh event statistics and with a very clean theoretical Standard Model prediction.
For these reasons, a NNLO (next-to-next-to leading order) prediction of the complete QED contributions and a NNLLO (next-to-next-to leading logarithmic order) prediction in the Standard Model are needed.
The virtual QED corrections at NNLO accuracy have been determined in a series of articles quite recently \cite{Bern:2000ie,Glover:2001ev,penin:2005kf,Bonciani:2004qt,Actis:2007gi,Becher:2007cu,Bonciani:2007eh,Actis:2007pn,Actis:2007fs}.
A complete evaluation of the photonic corrections covers additionally the real photon emission contributions and fermion pair production.

In this talk, we discuss one class of Feynman diagrams for real photon emission, namely radiative loop corrections,
\bea
e^+ + e^- \to e^+ + e^- + \gamma ,
\eea
which are contributing at NNLO to reaction (\ref{eq1}).
Their evaluation includes 5-point functions.
Usually, the scalar, vector, and tensor functions of this type will be reduced to simpler one-loop functions.
We also discuss an alternative approach, based on  Mellin-Barnes representations of Feynman parameter integrals.
%==========================================================================================
%==========================================================================================
\section{\label{secred}Reduction of 5-point functions}
%==========================================================================================
A critical point in an algebraic reduction of vector or higher tensor 5-point functions to scalar 2-point, 3-point, and 4-point functions (in four dimensions) 
is the appearance of inverse Gram determinants.
It is known that these inverse Gram determinants are spurious 
\cite{Bern:1994kr,Binoth:2005ff,Denner:2005nn}
and that they may be canceled out in the final analytical expressions.
For the approach proposed in \cite{Fleischer:1999hq}, we have demonstrated this cancellation quite recently.
Because that part of the presentation was decribed in some detail in other contexts 
\cite{Fleischer:2007ff,Fleischer:2007ph}, 
we don't repeat the material in these proceedings again. 

The focus will be on two questions:
\begin{itemize}
 \item
Is the MB-approach useful for the numerical evaluation of the finite parts of scalar, vector, and tensor 5-point functions?
\item
How to treat the infrared divergencies of these functions? 
\end{itemize}
%--
%=========================================================
%==========================================================================================
\section{\label{sec1}Mellin-Barnes representation for massive 5-point functions}
%==========================================================================================
The use of Mellin-Barnes (MB) integrals for the representation and evaluation of Feynman integrals has a long history, although a systematic use of it became possible quite recently.
The replacement of massive propagators by MB-integrals was proposed in \cite{Usyukina:1975yg} for a finite 3-point function.
It was worked out for one-loop $n$-point functions with arbitrary indices (powers of propagators) in $d=4-2\eps$ dimensions in \cite{Boos:1991rg,Davydychev:1991jt,Davydychev:1992cq}, where also
some of the related earlier literature is discussed, as well as the applicability to tensor integrals and to multi-loop problems.
The aim was a replacement of massive by massless propagators.
In \cite{Usyukina:1992jd}, the Feynman parameter representation (or $\alpha$-parameter representation, the difference plays no role here) was derived and then for the characteristic function of the diagrams an MB-representation was applied.
Along this line, a systematic approach to MB-presentations for divergent multi-loop integrals was derived and solved for non-trivial massless and massive cases \cite{Smirnov:1999gc,Smirnov:1999wz,Tausk:1999vh,Smirnov:2001cm,Smirnov:2004,Smirnov:2004ym,%
Heinrich:2004iq}.
Since software packages like 
AMBRE.m \cite{Gluza:2007rt} (in Mathematica, for the derivation of MB-representations),
MB.m \cite{Czakon:2005rk} (in Mathematica, for their analytical and series expansion in $\eps$), and XSUMMER  \cite{Moch:2005uc} (in FORM \cite{Vermaseren:2002rp}, for taking sums of their residues)
became publicly available, quite involved integrals may be treated, see e.g. \cite{Czakon:2006pa}.

Of course, such a complicated task like the evaluation of -- ideally -- arbitrary Feynman integrals will not be finally solved with using one or the other method.
In fact, already quite simple problems may be used to demonstrate the limitations of some approach.
We will study here, with MB-integrals, some one-loop functions of massive QED as occurring in Bhabha scattering, with focussing on the 5-point function shown in Figure \ref{5-point-sample}. 
  
\def\graphicsize{0.5}
\begin{figure}[bthp]
\begin{center}
\includegraphics[scale=\graphicsize]{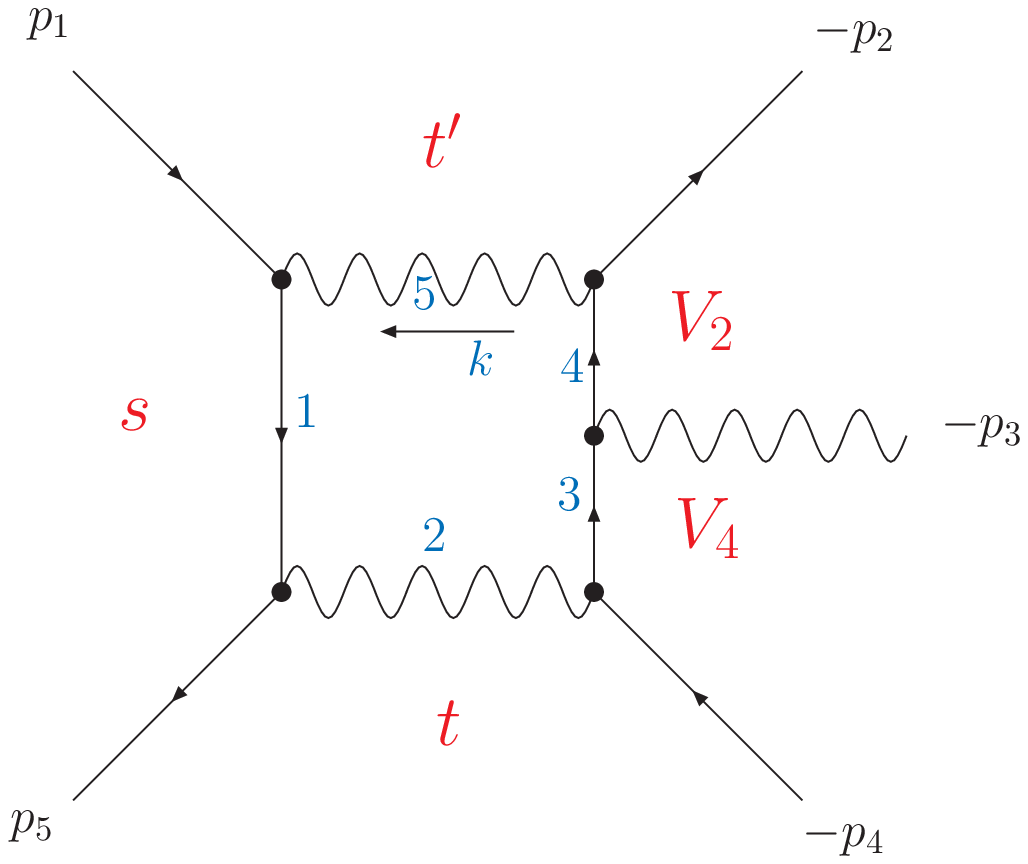}
\hspace*{1cm}
\includegraphics[scale=\graphicsize]{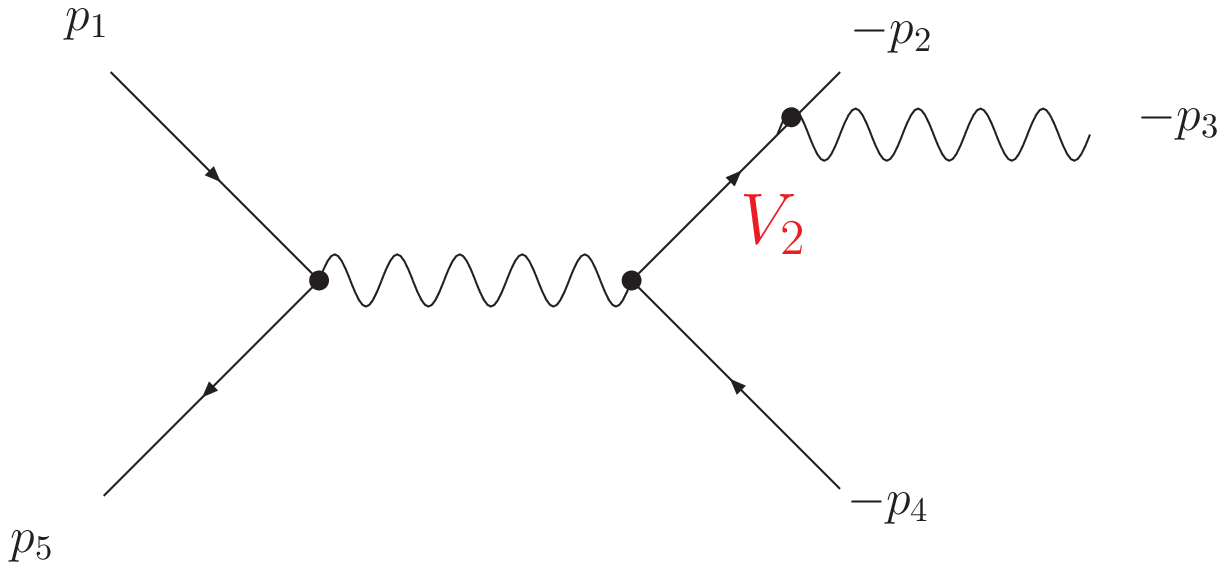}
\caption{A pentagon topology and a Born topology}
\label{5-point-sample}
\end{center}
\end{figure}

We define
\bea\label{i5defi}
%I_5^{\{0,\mu,\mu\nu,\mu\nu\rho\}} &=& 
I_5[A(q)] &=& 
e^{\eps\gamma_E} \int \frac{d^dq}{i\pi^{d/2}}
%\frac{\{1,q^{\mu},q^{\mu}q^{\nu},q^{\mu}q^{\nu}q^{\rho}\}}
\frac{A(q)}{d_1d_2d_3d_4d_5} ,
\eea
with the chords $Q_i$,
\bea
d_i = (q-Q_i)^2 - m_i^2.
\eea
This representation becomes unique after choosing one of the chords (and the direction of the loop momentum), e.g.:
\bea
Q_5^{\mu} = 0,~~~Q_1^{\mu} = p_1^{\mu} .
\eea
The numerator $A(q)$ contains the tensor structure, 
\bea
A(q) = \{ 1,q^{\mu},q^{\mu}q^{\nu},q^{\mu}q^{\nu}q^{\rho},\cdots \} ,
\eea
 or may be used to define pinched diagrams;
e.g. a shrinking of line 5 leads to a box diagram corresponding to 
\bea
I_5[d_5] = e^{\eps\gamma_E}
\int \frac{d^dq}{i\pi^{d/2}}
%\frac{\{1,q^{\mu},q^{\mu}q^{\nu},q^{\mu}q^{\nu}q^{\rho}\}}
\frac{1}{d_1d_2d_3d_4}.
\eea
%###################################################################
For details of the derivation of  Feynman parameter integrals we refer to any textbook on perturbative quantum field theory or to  \cite{Gluza:2007cp}.
A Feynman parameter representation for Fig. \ref{5-point-sample} is:
\bea
\label{bh-19}
I_5[A(q)]&=& - e^{\eps\gamma_E}\int_0^1 \prod_{j=1}^5 dx_j ~ 
\delta\left(1-\sum_{i=1}^5 x_i\right) 
\frac{\Gamma\left(3+\eps\right) }{{F(x)}^{3+\eps}} ~{B(q)},
\eea
with 
$B(1)=1, B(q^{\mu})=Q^{\mu}, B(q^{\mu}q^{\nu})= Q^{\mu}Q^{\nu} - \frac{1}{2}g^{\mu\nu}F(x)/(2+\eps)$, and $Q^{\mu} = \sum x_i Q_i^{\mu}$.
The diagram depends on five kinematical invariants and the $F$-form in (\ref{bh-19}) is:
\bea
\label{fform}
F(x) &=& m_e^2 (x_2+x_4+x_5)^2 + [-s] x_1 x_3 + [-{V_4}] x_3x_5 + [-t]x_2x_4 + [-t']x_2x_5 +[-\RE{V_2}]x_1x_4.
\nl
\eea
Henceforth, $m_e = 1$.
It is evident that the $F$-form cannot be made more compact.
After the introduction of seven subsequent Mellin-Barnes representations,
\bea
\frac{1}{[A(x)+Bx_ix_j]^R} &=& \frac{1}{2\pi i} \int_{\mathcal{C}} dz [A(x)]^{z} [Bx_ix_j]^{-R-z}
\frac{\Gamma(R+z)\Gamma(-z)}{\Gamma(R)}
,
\eea
one for each additive term in $F$,
we may perform the $x$-integrations using a generalization of the integral representation of the Beta function:
\bqa
\label{a1}
\int_0^1 \prod_{j=1}^N dx_j ~ x_j^{\alpha_j-1}
~ \delta\left(1-x_1 - \cdots -x_N\right)
&=&
\frac{\Gamma(\alpha_1) \cdots \Gamma(\alpha_N)}
{\Gamma\left(\alpha_1 + \cdots + \alpha_N\right)}.
\eqa
The final MB-integral may be easily derived using our Mathematica package AMBRE.m \cite{Gluza:2007rt} (see also example1.nb and example2.nb of the package).
The representation  is five-dimensional after twice applying Barnes' first lemma.
The integrals are well-defined on integration strips parallel to the imaginary axis, for a finite value of $\eps = d/2-2$.
After an analytical continuation in $\eps$, preferrably done by MB.m \cite{Czakon:2005rk}, one gets a sequence of finite, multi-dimensional  MB-integrals.
We performed these steps and met the following situation for the terms proportional to $1/\eps$ and $O(1)$:
\begin{itemize}
\item
scalar integrals: the  MB-integrals are up to three-fold;
\item
vector integrals: the  MB-integrals are up to three-fold;
\item
tensor integrals: the  MB-integrals are up to five-fold.
\end {itemize}
We performed some experimental calculations, but 
there is no need to go into more detail:
A numerical evaluation of these integrals, especially of the five-dimensional ones, in the Minkowskian region, is not competative to the old-fashioned numerical packages like FF \cite{vanOldenborgh:1991yc}, or LoopTools \cite{vanOldenborgh:1991yc,Hahn:1998yk2,Hahn:2006qw},
which rely on the preceding algebraic reduction of all the 5-point functions to well-known scalar 2- to 4-point functions.

For this reason, we restrict the discussion now to the infrared divergent parts only.
As is well-known, they have a lower dimensionality, and here the MB-presentations are well-suited.
As examples we will use the scalar and vector 5-point functions.
%==========================================================================================
\section{\label{sec3}Infrared singularities}
%==========================================================================================
Let us consider first the scalar function.
A set of five independent invariants may be read off from (\ref{fform}):
\bea
s&=&(p_1+p_5)^2, \nonumber \\ \nonumber
t&=&(p_4+p_5)^2,  \\\nonumber
t' &=& (p_1+p_2)^2,\\\nonumber
{V_2}&=&2 p_2 {p_3} ~~ \sim E_{3},  \\\nonumber
{V_4}&=&2 p_4 {p_3} ~~  \sim E_{3} ,
\label{quanties}
\eea
and the two massless propagators are 
 $ d_5=q^2$ and $d_2=(q+p_1+p_5)^2$.
In the IR-limit, where $E_3 \to 0$, it will be $t'\approx t$ and $0\leq V_2,V_4<<s,|t|$.
The leading IR-singularities are easily found algebraically from the following decomposition:
\bea
\frac{1}{d_1d_2d_3d_4d_5} &=& \frac{-1}{s}
\Biggl[ \frac{2(q-Q_5)(q-Q_2)}{d_1d_2d_3d_4d_5} 
+ \frac{1}{V_2}\left(\frac{2(q-Q_5)(q-Q_3)}{d_1d_3d_4d_5} - \frac{1}{d_1d_3d_4} - \frac{1}{d_1d_4d_5}
\right)
\nl
&&+~ \frac{1}{V_4}\left(\frac{2(q-Q_2)(q-Q_4)}{d_1d_2d_3d_4} - \frac{1}{d_1d_2d_3} - \frac{1}{d_1d_3d_4}
\right)
\Biggr].
\eea
The 4-point functions depend on the variables $(t,t',V_2)$ and $(t,t',V_4)$, respectively, and the leading IR-singularities 
 of $I_5$ trace back, by construction,  to the two IR-divergent 3-point functions:
\bea
\int \frac{d^d k}{d_1d_2d_3d_4d_5} =
\frac{1}{sV_2} \int \frac{d^d k}{d_1d_4d_5} +  \frac{1}{sV_4} \int\frac{d^d k}{d_1d_2d_3} + \cdots
%\nonumber \\ 
\label{two3p}
= \frac{1}{\eps}\left[ \frac{F(t')}{sV_2}+\frac{F(t)}{sV_4}\right] + \cdots
\nonumber \\ 
\eea
The integrals with numerators are constructed such that they are free of IR-singularities arising from the virtual photon lines.
It is of importance here to observe that the denominators $V_2$ and $V_4$ are proportional to the photon energy $E_{3}$ and thus give rise to additional IR-problems, stemming from the photon phase space integral over the squared sum of matrix elements; e.g.:
\bea\label{dvv}
\int \frac{d^3p_3}{2E_3} ~ \frac{A}{E_3} ~\frac{B(E_3)}{E_3} \to \int_0^{\omega} \frac{dE_3}{E_3} = \ln(E_3)|_0 ^{\omega} .
\eea 
Here, one term (${A}/{E_3}$) comes from the real photon emission Born diagram, and the other one ($B(E_3)/{E_3}$) from our pentagon diagram.  
After dimensional regularisation, this becomes evaluable and contributes also to the Laurent series in $\eps$.
We learn from (\ref{dvv}) that a complete treatment of the IR-problem includes a careful control of the subleading (and in $4$ dimensions non-integrable) terms like $1/V_i$ and $\ln(V_i)/V_j$.
This leads to phase space integrals with a behaviour like:
\bea
%notesIII87,IR-div-Integr.nb
\int_0^{\omega} \frac{dE_3}{E_3^{5-d}} \left(\frac{a}{\eps} + b \ln(E_3) +c\right)
&=&  - \frac{2a+b}{4\eps^2} -\frac{c-2a\ln(\omega)}{2\eps}
\nonumber \\
&& +~ c \ln(\omega) + \frac{1}{2}(2a+b) \ln^2(\omega) + O(\eps) .
\eea
Evidently, one separates with the 3-point functions in (\ref{two3p}) only a leading singularity, while we expect  expressions like
\bea
\int \frac{d^dk}{d_1d_2d_3d_4d_5} = \frac{A_2}{sV_2\eps} + \frac{A_4}{sV_4\eps} + \frac{B_2}{sV_2}\ln(V_2) 
 + \frac{B_4}{sV_4}\ln(V_4)+ \frac{C_2}{sV_2} +\frac{C_4}{sV_4} +\cdots
\eea
Subleading singuarities may arise from the $\eps$-finite 4- and 3-point functions with pre-factors $1/V_i$.

It is also evident that
the whole above discussion immediately transfers over to vector and tensor integrals.

Concentrating now on the IR-divergent parts,
we may safely assume  now the validity of the Born kinematics, including  
\bqa
t' = t,
\eqa
which is justified bcause of the vanishing photon momentum in this limit. 
This `eats' another MB-integration (in the $F$-form (\ref{fform}) one additive term vanishes), and the starting point of further discussions are four-dimensional MB-integrals.
For the scalar pentagon:
\bea
% radcor.nb of 27-09-2007
I_5 &=&
\frac{-e^{\eps\gamma_E}}{(2\pi i)^4}
\prod_{i=1}^4 \int\limits_{-i\infty+u_i}^{+i\infty+u_i} d z_i
 (-s)^{z_2} 
(-t)^{z_4} (-V_2)^{z_3} (-V_4)^{-3-\eps-z_1-z_2-z_3-z_4}
\frac{\prod\limits_{j=1}^{12}\Gamma_j}{\Gamma_0\Gamma_{13}\Gamma_{14}},
\nonumber\\
\eea
with a normalization 
$ \Gamma_{0}=\Gamma[-1-2\eps]$,
and the other $\Gamma$-functions are:
\bea\label{mbaux}
\Gamma_1&=&\Gamma[-z_1],
~~%\nl
\Gamma_2=\Gamma[-z_2],
~~%\nl
\Gamma_3=\Gamma[-z_3],
~~%\nl
\Gamma_4=\Gamma[1+z_3],
\nl
%\Gamma_5&=&\Gamma[d/2 -4 - r_1 - r_4],
 \Gamma_5&=&\Gamma[1+z_2+z_3],
~~%\nl
\Gamma_6=\Gamma[-z_4],
~~%\nl
\Gamma_7=\Gamma[1 +z_4],
~~%\\\nl
%\Gamma_8&=&\Gamma[d/2 -4  - r_1 - r_5],
 \Gamma_8=\Gamma[ -1-\eps - z_1 - z_2],
%%\nl
%\Gamma_9&=&\Gamma[d -8  - 2 r_2 - r_5 - r_6],
\nl
\Gamma_{9}&=&\Gamma[-2-\eps-z_1-z_2-z_3-z_4],
~~%\nl
\Gamma_{10}=\Gamma[-2-\eps-z_1-z_3-z_4],
%\nl
%\Gamma_{12}&=&\Gamma[2  + r_3 + 2 r_4 + r_6],
\nl
% corr due to SVT_comm.nb 16 02 2007
%\Gamma_{11}&=&\Gamma[-d/2 +5 +  r_1 + r_4 + r_5],
 \Gamma_{11}&=&\Gamma[-\eps+z_1-z_2+z_4],
~%\nl
\Gamma_{12}=\Gamma[3 +\eps+z_1+z_2+z_3+z_4],
\nonumber
\eea
and, in the denominator:
\bea
%\Gamma_{16}&=&\Gamma[d -8 - 2 r_2 - r_5 - r_6],
%\nl
%\Gamma_{17}&=&\Gamma[2 + r_3 + 2 r_4 + r_6],
%\nl
\Gamma_{13}&=&\Gamma[-1-\eps-z_1-z_2-z_4] ,
~~%\nl
\Gamma_{14}=\Gamma[-\eps-z_1-z_2+z_4] .
\eea
The $I_5$ is finite if all $\Gamma$-functions in the numerator  have positive real parts of the arguments; this may be fulfilled for finite $\eps$ (here we follow the method invented in \cite{Tausk:1999vh}):
\bea 
\eps = -\frac{3}{4} .
\eea
The real shifts $u_i$ of the integration strips $r_i$ may be chosen to be:
\bea
u_1 &=& -5/8, \nl
u_2 &=& -7/8 ,\nl
u_3 &=& -1/16, \nl
u_4 &=& -5/8,\nl
u_5 &=& -1/32.
\eea
%###################################################################
The further discussion of the scalar case is very similar to that of the QED vertex function given in \cite{Gluza:2007bd}, so we may concentrate here on the results for the IR-divergent part:
\bea
I_5^{IR} &=& I_5^{IR}(V_2) +I_5^{IR}(V_4),
\\\label{i5ir}
% qqqq
I_5^{IR} (V_i)&=& \frac{ I_{-1}^s(V_i)}{\eps} + I_0^s(V_i).
\eea
%With Mathematica or using Kalmykov et al., Huber and Maitre:
The explicit expressions for the inverse binomial sums solving  the MB-integrals are obtained by  applying the residue theorem (closing the integration contours to the left):
\bea\label{sm1}
\frac{I_{-1}(V_i)^s}{\eps} &=&
\frac{1}{2sV_i\eps} \sum_{n=0}^{\infty} 
\frac{(t)^n}{ \begin{pmatrix}2n\\n\end{pmatrix} (2n+1)} ,
\eea
with $I_{-1}^s$ being in accordance with (\ref{two3p}), and:
\bea
I_0(V_i)^s &=&\frac{1}{2sV_i}
 \sum_{n=0}^{\infty} 
\frac{(t)^n}{ \begin{pmatrix}2n\\n\end{pmatrix} (2n+1)}
\left[ -2\ln(-V_i) - 3 S_1(n) +2 S_1(2n+1)\right] ,
\eea
where we introduce the harmonic numbers $S_k(n) = \sum_{i=1}^{n}1/i^k$, and have to understand $\ln(-V_i) = \ln(V_i/s) + \ln[-(s+i\delta)/m_e^2]$.

The series may be summed up in terms of polylogarithmic functions with the aid of Table~1 of Appendix D of \cite{Davydychev:2003mv}:
\bea
\sum_{n=0}^\infty \frac{t^n}{
\begin{pmatrix}2n\\n\end{pmatrix}
%\left( 2n \atop n\right) 
(2n+1)}
&=&
\frac{y}{y^2-1} 2\ln (y),
\\
\sum_{n=0}^\infty \frac{t^n}
{%\left( 2n \atop n\right)
\begin{pmatrix}2n\\n\end{pmatrix}  (2n+1)}
S_1(n)&=&\frac{y}{y^2-1}
\left[ -4 \Li{2}{-y} - 4 \ln (y) \ln (1+y)\right.
\nonumber 
\\ &&
+\left.\ln^2 (y) - 2\zeta_2 \right],
\label{S1_y}
\\
\sum_{n=0}^\infty \frac{t^n}
{\begin{pmatrix}2n\\n\end{pmatrix} %\left( 2n \atop n\right)
 (2n+1)}S_1(2n+1)&=&\frac{y}{y^2-1} \biggl[
2\Li{2}{y} - 4 \Li{2}{-y} - 4 \ln (y) \ln(1+y)
\nonumber \\ &&
+ 2\ln (y) \ln (1-y)
+ \frac{1}{2} \ln^2 (y)
- 4 \zeta_2   \biggr],
\eea
with
\bea
y~\equiv~y(t) &=& \frac{\sqrt{1-4/t}-1}{\sqrt{1-4/t}+1}.
\eea

For the vector and higher tensor 5-point functions one gets quite similar results.
The IR-divergent pieces arise only from those contributions, which are proportional to the chords $Q_2$ and $Q_5$  of the massless internal lines (one of them is set to zero here, $Q_5=0$):
\bea
I_5^{IR}[q^{\mu}] &=& Q_2^{\mu}\left( \frac{ I_{-1}^v(V_2,V_4)}{\eps} + I_0^v(V_2,V_4)\right).
\label{i5ira}
\eea
The MB-integrals introduced in (\ref{bh-19}) will not get modified by the additional factors $B(q^{\mu})$ etc., but the subsequent $x$-integrations will.
For the vector integrals, we obtain:
\bea
I_5[q^{\mu}]|_{t'=t} &=& \sum_{i=1}^5 Q_i^{\mu} I_5(i),
\eea
and
\bea
I_5(2) 
&=& 
\frac{-e^{\eps\gamma_E}}{(2\pi i)^4}
\prod_{i=1}^4 \int\limits_{-i\infty+u_i}^{+i\infty+u_i} d z_i
 (-s)^{z_2} 
(-t)^{z_4} (-V_2)^{z_3} (-V_4)^{-3-\eps-z_1-z_2-z_3-z_4}
\frac{\prod\limits_{j=1}^{12}\Gamma_j^v}{\Gamma_0^v\Gamma_{13}^v\Gamma_{14}^v},
\nonumber\\
\eea
where it is $\Gamma_j^v = \Gamma_j$ with two exceptions:
\bea
\Gamma_{10}^v &=& \Gamma[-1 - \eps - z_1 - z_3 - z_4],
\nonumber \\
\Gamma_0^v &=& \Gamma[-2\eps].
\eea
%vector = 
%------------------------------11.12.2007 from radcor-vectors.nb, forprint = ...
%-((Q2*(-s)^z2*(-t)^z4*(-V2)^z3*
%   (-V4)^(-3 - eps - z1 - z2 - z3 - z4)*Z*
%Gamma1[-z1]*
%Gamma2[-z2]*
%Gamma3[-z3]*
%Gamma4[1 + z3]*
%Gamma5[1 + z2 + z3]*
%Gamma6[-z4]*
%Gamma7[1 + z4]
%Gamma8[-1 - eps - z1 - z2]*
%Gamma9[-2 - eps - z1 - z2 - z3 - z4]*
%%was Gamma10[-2 - eps - z1 - z3 - z4]
%Gamma10a[-1 - eps - z1 - z3 - z4]*
%Gamma11[-eps + z1 - z2 + z4]*
%Gamma12[3 + eps + z1 + z2 + z3 + z4])
%/
%%
%Gamma0a[-2*eps]*
%Gamma13[-1 - eps - z1 - z2 - z4]*
%Gamma14[-eps - z1 - z2 + z4]))
%\eea
%

After similar manipulations as described above, we obtain finally for the IR-divergent part of the vector pentagon (and, not discussed at all, the tensor pentagon):
\bea
I_5^{IR}[q^{\mu}] &=& Q_2^{\mu}~I_5^{IR}(V_4)+ Q_5^{\mu}~I_5^{IR}(V_2),
\eea
and 
\bea
I_5^{IR}[q^{\mu\nu}] &=& Q_2^{\mu}Q_2^{\nu}~I_5^{IR}(V_4)+Q_5^{\mu}Q_5^{\nu}~I_5^{IR}(V_2).
\eea
In the above derivations, we chose arbitrarily $Q_5=0$.
The leading and non-leading IR-divergent parts of the tensor functions are contained in those terms of the tensor decomposition, which are proportional to the chords of the massless internal lines, and they agree with the corresponding scalar functions. 

\bigskip

\emph{In conclusion}, 
we have demonstrated, by analysing the loop functions without squaring matrix elements, that IR divergencies of scalar and tensor one-loop pentagon diagrams can be treated
in a systematic, efficient  way by using Mellin-Barnes representations.
The leading singularities of the $\epsilon$ expansion of MB-integrals are
obtained straightforwardly and have the same IR-structure as the vertex functions obtained by quenching.
Both the leading and non-leading singular parts (the latter being kinematical end point singularities) can be expressed by a few  well-known inverse binomial sums  or, equivalently, polylogarithmic functions.
The IR-structure of vector and tensor functions is completely reducible to that of the scalar function.

%==========================================================================================
%==========================================================================================
%==========================================================================================
\section*{Acknowledgements}
We would like to thank J. Fleischer and K. Kajda for a fruitful cooperation
related to the presented material.
\\
Work supported in part by 
SFB/TRR 9 of DFG
and by
MRTN-CT-2006-035505 ``HEPTOOLS'' and 
MRTN-CT-2006-035482 ``FLAVIAnet''.
%==========================================================================================
%==========================================================================================

% ****************************************************************************
% BIBLIOGRAPHY AREA
% ****************************************************************************

\begin{footnotesize}
% IF YOU DO NOT USE BIBTEX, USE THE FOLLOWING SAMPLE SCHEME FOR THE REFERENCES
% ----------------------------------------------------------------------------
%\begin{thebibliography}{99}
% Please replace the numbers for   contribId   and   sessionId
% in the following URL. You can get this information by going to 
% http://indico.cern.ch/confAuthorIndex.py?confId=9499
% and search for your contribution and click on the title
% Be aware: '&amp;' must be replaced by simple '&' as in example below
%%%%\bibitem{url} Slides: \\ 
%%%%\verb$http://ilcagenda.linearcollider.org/contributionDisplay.py?contribId=42&sessionId=8&confId=1296$

%.bbl follows
\providecommand{\href}[2]{#2}
% ----------------------------------------------------------------------------

% IF YOU USE BIBTEX,
% - DELETE THE TEXT BETWEEN THE TWO ABOVE DASHED LINES
% - UNCOMMENT THE NEXT TWO LINES AND REPLACE 'Name_Of_Your_BibFile'

%==========================================
\providecommand{\href}[2]{#2}

%\bibliographystyle{%
%%%%utphys_spires_tit.bst%
%utphys_spires%
%}
%\bibliography{2loops.bib}
%\include{riemann_tord_ew.bbl}

%==========================================

\end{footnotesize}

% ****************************************************************************
% END OF BIBLIOGRAPHY AREA
% ****************************************************************************

\end{document}